# $T\bar{T}$ deformation on non-Hermitian two coupled SYK model

Chenhao Zhang and Wenhe Cai[*]

*Department of Physics, Shanghai University, Shanghai, 200444, China*

We investigate $T\bar{T}$ deformation on non-Hermitian coupled Sachdev-Ye-Kitaev (SYK) model and the holographic picture. The relationship between ground state and thermofield double state is preserved under the $T\bar{T}$ deformation. We prove that $T\bar{T}$ deformed theory provides a reparametrization in large $N$ limit. We numerically calculate the deformation effects on the correlation function in imaginary time and the free energy. The thermodynamic phase structures show the equivalence of the wormhole-black hole picture between non-Hermitian model and Hermitian model still holds under $T\bar{T}$ deformation. The correlation function in Lorentz time, revival dynamic, and the deformed holography are also evaluated.

DOI: 10.1103/PhysRevD.110.086016

## I. INTRODUCTION

SYK model consists of Majorana fermions with Gaussian random coupling [1–4]. It can be numerically solved in large $N$ limit. There is a low-energy emergent conformal symmetry [5]. This model is dual to nearly anti–de Sitter (NAdS) black hole. An eternal traversable wormhole is constructed by Maldacena and Qi [6], which is identical to two coupled SYK model. In previous works [7], the causality of wormhole is discussed with the quench dynamics. A first order transition appears in the replica entropy, and the holographic information retrieval on boundaries are also investigated [8]. The finite temperature spectral function of coupled SYK consist of quasiparticles time [9]. The SYK tunneling spectroscopy, transition probability, and the corresponding black hole-wormhole geometry were discussed in [10].

Parts of non-Hermitian Hamiltonians preserve the (PT) symmetry with real spectra [11,12], systems with non-Hermiticity are gaining attention. The conditions of PT symmetry are weaker than Hermiticity; the investigation on non-Hermitian systems expand the study of quantum mechanics with unusual physical properties. For example, Ising approximation describes well the strong coupling limit for a non-Hermitian system, and the study of double scaling limit has been discussed [13]. The non-Hermitian Hamiltonian also exhibits bound states and binding energy with interesting properties in a weak coupling limit [14]. Single-side random SYK coupling is also redefined as non-Hermiticity, and the spectrums is both complex and real [15,16].

However, the interaction term between left and right side plays an important role in the eternal traversable wormhole (see [10] for a sample of complex MQ model). Therefore, we can directly incorporate a non-Hermitian bootstrap into the interaction term [17], and this model has a real spectrum. The non-Hermiticity in SYK model and interaction term does not violate the thermal structure or zero energy condition. Additionally, it preserves causality of travelable wormholes between boundaries. It provides thermodynamic structure equivalent to a Hermitian two coupled model. The $T\bar{T}$ deformation on the original uncoupled $SYK_4$ and $SYK_2$ model has been studied [18,19]. $T\bar{T}$ deformation effectively couples two local $SYK_4$ systems.

$T\bar{T}$ deformation [20–22] is an irrelevant deformation from IR to UV, with well-defined IR behaviors and a controllable UV limit. In two-dimensional theory, we can construct a composite operator $T\bar{T}$ via its energy-momentum tensor and further obtain the Laplace relation $L(\lambda)$ with an additional deformation parameter $\lambda$. A deformation theory is defined with $T\bar{T}$ operator. The flow equation is

$$\frac{d\mathcal{L}^\lambda}{d\lambda} = \frac{1}{2}\epsilon^{\mu\nu}\epsilon^{mn}T^\lambda_{\mu\nu}T^\lambda_{mn}. \qquad (1)$$

$T\bar{T}$ deformation can be extended to a one-dimensional theory, while $T\bar{T}$ flow equations are diagonalized to a 1D energy scalar [19]. This property provides an application to a $T\bar{T}$ deformed $AdS_2/CFT_1$ [23,24], and some additional physical quantities can be studied [25–27]. The $T\bar{T}$ deformation breaks conformal fields but not the complete holographic duality theory [28–31]. This nonlocal flow equation is similar to an effective bosonic string [32,33] and also induces correlation between different quantum systems.

[*]Contact author: whcai@shu.edu.cn

$T\bar{T}$ deformation also preserves integrability and maximal entanglement [34,35]. The $T\bar{T}$ operator is also a universal well-defined operator in both IR and UV. The original integrability and holography stay manifest under deformation [36–38]. Moreover, the $T\bar{T}$ deformation on AdS/CFT could also be explained in several perspectives [39–42].

In this work, the first motivation is that the $T\bar{T}$ deformation provides us a new kind of physics. Specially, the $T\bar{T}$ deformed SYK and the black holes in $NAdS$ spacetime have abundant interesting features. It is natural to further investigate the $T\bar{T}$ deformation on the coupled SYK model. It would be useful to investigate the deformed phase structures and the holography wormhole beyond common NAdS/NCFT.

The second motivation arises with non-Hermiticity. The non-Hermitian coupled SYK model is defined as pseudo-Hermitian, and this non-Hermiticity is not obviously distinguishable. We expect $T\bar{T}$ deformation on these theories could be influenced by non-Hermiticity, both dynamically and thermally. And we will discuss whether the thermal structure of Hermitian and non-Hermitian coupled SYK is distinguished under $T\bar{T}$ deformation.

The main goal in this paper is to investigate whether the wormhole-black hole picture is robust not only in the Hermitian two coupled SYK model, but also in a non-Hermitian two coupled SYK model under $T\bar{T}$ deformation.

The paper is organized as follows. In Sec. II, we propose the $T\bar{T}$ deformed coupled SYK models in Hermitian and non-Hermitian cases. In Sec. III, we start with the Thermofield double state under $T\bar{T}$ deformation. Then, we match the overlap between the corresponding TFD and interaction model to evaluate the transmission property of wormhole before or after $T\bar{T}$ deformation. In Sec. IV, we obtain the effective solution for large $N$ limit action. Then we evaluate the Green's function in an imaginary time and thermal phase structure. In Sec. V, we study revival dynamics of coupled cSYK model. The correlation functions in a Lorentz time of black hole and wormhole situation have been obtained numerically. In Sec. VI, we perform a $T\bar{T}$ deformation on NAdS spacetime and its reparametrized Schwarzian action to investigate the effects of $T\bar{T}$ deformation on holography. Section VII is the conclusion and discussion.

## II. THE MODEL

We consider a non-Hermitian two coupled SYK model [17],

$$H_0 = -\sum_{ijkl}^{N} J_{ijkl} \sum_{A=R,L} (2C_i^{A\dagger} C_j^{A\dagger} C_k^A C_l^A + 4C_i^{A\dagger} C_j^A C_k^{A\dagger} C_l^A) + H_{\text{int}},$$
$$H_{\text{int}} = \sum_i (-\kappa C_i^{L\dagger} C_i^L - \kappa C_i^{R\dagger} C_i^R + i\mu e^{-2\alpha} C_i^{L\dagger} C_i^R - i\mu e^{2\alpha} C_i^{R\dagger} C_i^L), \qquad (2)$$

with real $\kappa$, $\mu$, $\alpha$. Here, $H_L$ and $H_R$ refers to the single side SYK model with complex fermions [43,44]. The model is pseudo-Hermitian and reduces to a pseudocomplex SYK model when $\alpha = 0$, and exhibits a conservation charge $Q = \sum_i (C_i^\dagger C_i - \frac{1}{2})$. The parameter $\kappa$ has the same effects as parameter $\mu$ does [44]. The free energy is not influenced by non-Hermiticity [17]. The fermions and random coupling $J$ satisfies

$$J_{ijkl} = -J_{jikl} = -J_{ijlk} = J_{klij}, \langle J_{ijkl}\rangle = 0, \quad \langle J_{ijkl}^2\rangle = \frac{J^2}{8N^3},$$
$$\psi_{2i-1}^L = e^\alpha C_i^L + e^{-\alpha} C_i^{L\dagger}, \quad \psi_{2i-1}^R = e^{-\alpha} C_i^R + e^\alpha C_i^{R\dagger}, \qquad (3)$$

where Dirac fermions $C_i^\dagger$ and $C_i$ are generated from Majorana fermions $\psi_i$ in a MQ model [5], with non-Hermitian self-similarity transformation in (3). For simplicity, the components $\psi_{2i}^L = ie^\alpha C_i^L - ie^{-\alpha} C_i^{L\dagger}$ and $\psi_{2i}^R = ie^{-\alpha} C_i^R - ie^\alpha C_i^{R\dagger}$ are ignored under statistical symmetry in [17] when we consider creation and annihilation operators. And terms such as $\psi_i \partial \psi_i$ exist and their inclusion does not compromise generality. Since $T\bar{T}$ deformation is triggered by 2D energy-momentum tensor, we can consider a $T\bar{T}$ operator by energy-momentum tensor,

$$\mathcal{L}^{(\lambda+\delta\lambda)} = \mathcal{L}^{(\lambda)} - \delta\lambda T\bar{T}^{(\lambda)},$$
$$\frac{d\mathcal{L}^{(\lambda)}}{d\lambda} = \det T_{\mu\nu}^{(\lambda)}. \qquad (4)$$

Moreover, $T\bar{T}$ deformation is still well defined in 1D theory, since the tensor operator and flow equation are diagonalized by $T = H$, and we have

$$\frac{\partial S}{\partial \lambda} = \int d\tau \frac{T^2}{1/2 - 2\lambda T}. \qquad (5)$$

This equation could be solved in a simple Nambu-Goto form [20],

$$H(\lambda) = f(H_0, \lambda) = \frac{1}{4\lambda}(1 - \sqrt{1 - 8\lambda(H_0 - E_0)}),$$

$$L_E(H_0, \lambda) = \sum_i \psi_i \partial_\tau \psi_i + \frac{1}{4\lambda}(1 - \sqrt{1 - 8\lambda(H_0 - E_0)}). \quad (6)$$

$T\bar{T}$ operator $T$ is equivalent to the Hamiltonian. In thermodynamics, $T\bar{T}$ deformation is equal to a shifting on Hamiltonian with a nonlocal effect. Parameter $E_0$ is an arbitrary constant mathematically, which could be determined by a physical fixed point. And we will see that parameter $E_0$ could be set as 0 in large $N$ effective action in Sec. IV.

To employ the disorder-averaged Nambu-Goto action, we can apply a trick to linearize the Hamiltonian by introducing fermions $\psi_i$ and auxiliary field $\zeta$ [18,24],

$$S_E(H_0, \lambda, \zeta) = \int d\tau \left( \frac{1}{2} \sum_i \sum_{A=L,R} \psi_i^A \partial_\tau \psi_i^A - \frac{\zeta}{8\lambda}(1 - \zeta^{-1})^2 + \zeta H_0 - \zeta E_0 \right), \quad (7)$$

with the constraint,

$$H(\lambda) + H_0 - E_0 - 2\lambda H(\lambda)^2 = 0. \quad (8)$$

$T\bar{T}$ deformation on non-Hermitian two coupled SYK model can be written as

$$S_E(\lambda, \zeta) = \int d\tau \left( \sum_{A=R,L} \sum_i C_i^A \partial_\tau C_i^A - \frac{\zeta}{8\lambda}(1 - \zeta^{-1})^2 - \zeta E_0 \right) - \frac{36}{4} J^2 N \int d\tau d\tau' \sum_{A,B=R,L} \zeta(\tau)\zeta(\tau') G_{AB}^2(\tau, \tau') G_{BA}^2(\tau', \tau)$$

$$+ N \int d\tau d\tau' \zeta(\tau)\zeta(\tau')(i\mu e^{2\alpha} G_{LR}(\tau, \tau')\delta(\tau' - \tau) - i\mu e^{-2\alpha} G_{RL}(\tau', \tau)\delta(\tau - \tau'))$$

$$- N \int d\tau d\tau' \zeta(\tau)\zeta(\tau')(\kappa G_{LL}(\tau, \tau')\delta(\tau' - \tau) + \kappa G_{RR}(\tau, \tau')\delta(\tau' - \tau)). \quad (9)$$

The Hamiltonian is written with Green's function,

$$G_{AB}(\tau, \tau') = \frac{1}{N} \sum_i \langle \psi_n^l | \mathcal{T} C_i^{A\dagger}(\tau) C_i^B(\tau') | \psi_n^r \rangle, \quad (10)$$

where bra $\langle \psi_n^l |$ and ket $|\psi_n^r \rangle$ are biorthogonal states and not symmetry due to non-Hermitian coupling,

$$H|\psi_n^r\rangle = E_n|\psi_n^r\rangle, \quad H^\dagger|\psi_n^l\rangle = E_n^*|\psi_n^l\rangle$$

$$\sum_i |\psi_i^l\rangle\langle\psi_i^r| = I, \quad \langle\psi_m^l|\psi_n^r\rangle = \delta_{mn}. \quad (11)$$

The auxiliary field $\zeta$ is a solution to Eq. (9),

$$\zeta^{-2} = 1 + a - 8\lambda H_0, \quad (12)$$

where $a = 8\lambda E_0$. The $\zeta$ should be consistent with time translation symmetry and time dependence has been integrated out. The third term denotes the bare Hamiltonian $H_0$. We introduce effective deformation parameter $b(\lambda)$, which is proportional to the bare deformation parameter $\lambda$, to absorb the additional constant in integrals and Fourier transformation. With

$$8\lambda H_0 = -\frac{36}{4}(8\lambda) J^2 N \int d\tau d\tau' \sum_{A,B=R,L} G_{AB}^2(\tau, \tau') G_{BA}^2(\tau', \tau) + 8\lambda N \int d\tau d\tau' (i\mu e^{2\alpha} G_{LR}(\tau, \tau')\delta(\tau; -\tau)$$

$$- i\mu e^{-2\alpha} G_{RL}(\tau', \tau)\delta(\tau - \tau')) - 8\lambda N \int d\tau d\tau' (\kappa G_{LL}(\tau, \tau')\delta(\tau' - \tau) + \kappa G_{RR}(\tau, \tau')\delta(\tau' - \tau)).$$

$$= -bT \sum_{A,B=R,L} \sum_{\omega_n} \left( G_{AB}^2(i\omega_n) G_{BA}^2(i\omega_n) - \frac{1}{9} i\mu(-e^{2\alpha} G_{LR}(i\omega_n) + e^{-2\alpha} G_{RL}(i\omega_n)) + \frac{1}{9} \kappa(G_{LL}(i\omega_n) + G_{RR}(i\omega_n)) \right). \quad (13)$$

In Eq. (4), $T\bar{T}$ deformation also provides additional entanglement with chaotic behaviors [19,24]. According to statistical definition of $T\bar{T}$ operator, the energy-momentum flow should converge to 0 without a vertical excitation in the low temperature limit. For example, the Hamiltonian in Eq. (13) could also be statically approximated by 0 when $T \to 0$. The $T\bar{T}$ deformation can be equally transformed into a constant shift of ground state energy for the additional arbitrary parameter in Nambu-Goto solution. For simplicity, this constant $E_0$ can be set to 0 without breaking any physical observation. And it means an arbitrary deformation fix point has been moved into the low temperature limit. When we consider a non-Hermitian model, $T\bar{T}$ deformation expands the flow equation into complex plane. Since coupled SYK model (2) is pseudo-Hermitian, the non-Hermitian $T\bar{T}$ deformation [45,46] reduces to a Hermitian version.

### III. THERMOFIELD DOUBLE STATE

The thermofield double (TFD) state successfully represents some indispensable structures of quantum wormhole. It enables an information transformation ability between two identical SYK models (such as entropy and transition time were addressed in [8]). Then it is natural to ask whether these features are influenced by $T\bar{T}$ deformation.

Since the thermal deformation could be represented by mapping the Hamiltonian in a certain way, the $T\bar{T}$ deformed thermofield double states are also generated from the undeformed one,

$$|TFD_\lambda(\beta)\rangle = \exp(-\beta f(H_L + H_R, \lambda)/2)|TFD_0(0)\rangle. \quad (14)$$

Here, $\beta = 1/T$. The definition of TFD state is irrelevant to the coupled part, so the $T\bar{T}$ deformed TFD should be performed independently. The TFD state is a special quantum algebraic system, which is related to thermal structure and the Hamiltonian in left and right copy Hilbert spaces,

$$|\text{TFD}_0(0)\rangle = \frac{1}{2^{N/2}} \sum_q \sum_{n_q} |n_q\rangle_L \otimes \left(e^{\frac{-in_q\pi\Gamma}{4}} e^{iq(\frac{\pi}{2}-\phi)}\right) P|n_q\rangle_R, \quad (15)$$

where $q$ represents eigenvalues of Fermionic charge in a complex variable, and $\sum_{n_q} |n_q\rangle$ are set of states with charge $q$. The left and right conjugate Hilbert space in TFD should satisfy the fermion parity condition,

$$\Gamma = (-1)^{q+\frac{N}{2}}, \qquad P^{-1} C_i P = \eta C_i^\dagger, \quad (16)$$

and the original $\text{TFD}_0$ state is maximally entangled. The fermionic partial $\eta = \pm 1$. The operator $P$ is introduced to represent the parity conjugate on left and right sides, which brings a special fermion symmetry in TFD Hilbert space. $\phi$ is an additional arbitrary phase-evolving degree of freedom between two SYK models, which is related to the conserved charge of a cSYK model.

When we set coupling $\mu$ close to infinite, equations in a strong coupling limit are obtained. Left or right SYK Hamiltonians have been ignored and a deformed Hamiltonian of the entire system turns out to be

$$f(H) = f(H_{\text{int}}). \quad (17)$$

Diagonal element of $\text{TFD}_\beta = 0$ has been introduced in undeformed system,

$$\langle \text{TFD}_{\beta=0}|H_{\text{int}}|\text{TFD}_{\beta=0}\rangle = \frac{1}{2^N} \sum_{qq'} \sum_{mn} \langle \bar{n}_{-q'}|_R \otimes \langle n_{q'}|_L \left( i\mu \sum_i (e^{-2\alpha} C_i^{L\dagger} C_i^R - e^{2\alpha} C_i^{R\dagger} C_i^L) \right) |m_q\rangle_L \otimes |\bar{m}_{-q}\rangle_R$$

$$= \frac{1}{2^N} \sum_{qq'} \sum_{nm} \sum_i \left( i\mu e^{i\pi(q+\frac{N}{2})} e^{-2\alpha} \langle n_{q'}|C_i^{L\dagger}|m_q\rangle_L \langle \bar{n}_{-q'}|C_i^R|\bar{m}_{-q}\rangle_R \right.$$

$$\left. - i\mu e^{i\pi(q'+\frac{N}{2})} e^{2\alpha} \langle \bar{n}_{-q'}|C_i^{R\dagger}|\bar{m}_{-q}\rangle_R \langle n_{q'}|C_i^L|m_q\rangle_L \right)$$

$$= \sum_{qq'} \sum_{nm} \sum_i (i\mu e^{2\alpha} - i\mu e^{-2\alpha}) e^{-i\phi}/2^N \langle n_{q'}|C_i^{L\dagger}|m_q\rangle_L \langle m_q|C_i^L|n_{q'}\rangle_L$$

$$= -\mu(e^{-2\alpha} + e^{2\alpha})/2^N \sum_q \sum_n \sum_i \langle n_{q'}|C_i^{L\dagger} C_i^L|n_q\rangle_L = -\frac{\mu N}{2}(e^{-2\alpha} + e^{2\alpha}). \quad (18)$$

Here, the matrix elements in Hilbert space are generated by an operators-represented Hamiltonian,

$$\langle \bar{n}_{-q'}|C_i^R|\bar{m}'_q\rangle_R = -e^{i\pi(q+\frac{N}{2})} e^{-i\phi} \langle m_q|C_i^L|n_{q'}\rangle_L,$$
$$\langle \bar{n}_{-q'}|C_i^{R\dagger}|\bar{m}'_q\rangle_R = -e^{i\pi(q'+\frac{N}{2})} e^{-i\phi} \langle m_q|C_i^{L\dagger}|n_{q'}\rangle_L. \quad (19)$$

It is natural to define the $T\bar{T}$ deformation has no influence on zero-temperature theory. Similarly, we involve the deformed disorder average in the strong coupling limit. $T\bar{T}$ deformation can be computed by the mapping Hamiltonian with the deformed Nambu-Goto form (6),

$$\langle f(H)\rangle = \langle f(H_{\text{int}})\rangle = f(\langle H_{\text{int}}\rangle). \quad (20)$$

Likewise, we can also use the same process in a weak coupling limit by ignoring the interaction term; a reduced Hamiltonian is obtained as

$$f(H) = f(H_L + H_R). \tag{21}$$

Similar to strong coupling limit, a finite deformation seems to be less important when $\beta$ approaches infinity,

$$\langle \text{TFD}_{\beta \to \infty} | H | \text{TFD}_{\beta \to \infty} \rangle = \prod_{i=1}^{N} \langle \bar{0}|_{R,i} \langle 0|_{L,i} \sum_{ijkl} J_{ijkl} \sum_{a=L,R} (2C_i^{A\dagger} C_j^{A\dagger} C_k^A C_l^A + 4C_i^{A\dagger} C_j^A C_k^{A\dagger} C_l^A) |0\rangle_{L,i} |\bar{0}\rangle_{R,i} \tag{22}$$

where $|0\rangle$ and $|\bar{0}\rangle$ are vacuum states in partial Hilbert spaces. Obviously, this result is independent of the interaction coupling $\mu$, $\text{TFD}_{\beta=\infty}$ vacuum, and the SYK eigenstate.

However, this result on overlap is no longer fixed in a more general $\beta$ condition. Numerically, we obtain an approximate overlap $\langle \text{TFD}_\lambda(\beta)|G\rangle$ between an arbitrary parameter $\beta$ and the ground states with the best-fixing coupling constant $\mu$. The ground states with a non-Hermitian bootstrap on the interaction term are proposed as

$$|G\rangle = \prod_{i=1}^{N} \frac{1}{\sqrt{1+e^{4\alpha}}} (|1\rangle_{L,i}|0\rangle_{R,i} \pm ie^{2\alpha}|0\rangle_{L,i}|1\rangle_{R,i}). \tag{23}$$

Here, we define the excited states $|1\rangle_{A,i} = C_i^{A\dagger}|0\rangle_{A,i}$.

Overlap between the ground state of $H_{\text{int}}$ and TFD has an important role in exploring the entanglement property and transformation between two Cartesian product Hilbert space. Since we simply introduce the non-Hermiticity into operators (2) without changing the Hilbert space or overlap. We can also consider the $T\bar{T}$ deformed non-Hermitian overlap between the deformed TFD state, which is generated with a thermally deformed copied Hamiltonian and its best-fixing valid deformed ground states,

$$\langle \text{TFD}_\lambda(\beta)|G\rangle = \langle \text{TFD}_0(\beta)| \frac{\sqrt{Z_0(\beta)}}{\sqrt{Z_\lambda(\beta)}} \exp\left(-\beta(H_L + H_R - f(H_L + H_R, \lambda))/2\right)|G\rangle. \tag{24}$$

Conventionally, ground states in an overlap calculation could be generated by a simple infinite evolution, while the Hamiltonian has been substituted by a deformed one,

$$|G(\tau)\rangle = \exp\left(-\tau(H_L + H_R + H_{\text{int}})\right)|G\rangle. \tag{25}$$

Since the TFD states are independent of the interaction $\mu$, we can simply choose another ground state with a different coupling $\mu$ to remove this effect,

$$\langle \text{TFD}_\lambda(\beta)|G\rangle = \langle \text{TFD}_0(\beta)| \frac{\sqrt{Z_0(\beta)}}{\sqrt{Z_\lambda(\beta)}} \exp\left(-\beta(H_L + H_R - f(H_L + H_R, \lambda))/2 - \tau(H_L + H_R + H_{\text{int}})\right)|0\rangle$$
$$= \langle \text{TFD}_0(\beta)|G'\rangle. \tag{26}$$

Coefficient $\sqrt{Z_\lambda(\beta)}$ is introduced to normal the partition function $Z$. Because the overlap equations have 2 degrees of freedom, the evolution of TFD states includes a finite deformation function $f$ and a decoupled SYK Hamiltonian $H$, while the ground states includes the interaction coupling $\mu$. In principle, we can easily absorb TFD deformation by shifting the parameter $\mu$ without altering physical effects. And the deformed TFD is equal to the undeformed theory with different coupling without causing any additional divergence.

## IV. THERMAL PHASE STRUCTURE

In this section, we first consider the effective action of the non-Hermitian coupled SYK model under the $T\bar{T}$ deformation in the large $N$ limit,

$$\frac{S_E}{N} = -\log \det(\sigma_{AB} - \Sigma_{AB}) - \int d\tau d\tau' \sum_{AB} [\Sigma_{AB}(\tau, \tau')G_{BA}(\tau', \tau) + 9J'^2 G_{AB}^2(\tau, \tau')G_{BA}^2(\tau', \tau)],$$

$$\sigma_{AB} = \begin{pmatrix} \partial_\tau - \kappa' & i\mu' e^{-2\alpha} \\ -i\mu' e^{2\alpha} & \partial_\tau - \kappa' \end{pmatrix}, \Sigma_{AB}(\tau, \tau') = -36J'^2 G_{AB}^2(\tau, \tau')G_{BA}(\tau', \tau), \tag{27}$$

where

$$J'^2 = J^2 \zeta(\tau)\zeta(\tau'), \quad \mu' = \mu\zeta(\tau)\zeta(\tau'), \quad \kappa' = \kappa\zeta(\tau)\zeta(\tau'). \tag{28}$$

Because the auxiliary field $\zeta$ is dimensionless, the $T\bar{T}$ deformed model in thermal limit behaves exactly like a shifting coupling constant. We can rescale $J^2, \kappa, \mu$ by $\zeta(\tau)\zeta(\tau')$. After Fourier transformation, the auxiliary field $\zeta$ in Eq. (12) becomes

$$\zeta^{-2} = 1 + bT \sum_{\omega_n} \left( \sum_{AB} G_{AB}^2(i\omega_n) G_{BA}^2(i\omega_n) - \frac{1}{9} i\mu(-e^{2\alpha} G_{LR}(i\omega_n) + e^{-2\alpha} G_{RL}(i\omega_n)) + \frac{1}{9} \kappa(G_{LL}(i\omega_n) + G_{RR}(i\omega_n)) \right). \tag{29}$$

The saddle point equations with Fourier transformation is derived from Eq. (27),

$$G_{AB}(i\omega_n, \alpha) = \frac{(-i\omega_n - \kappa')\delta_{AB} - \Sigma_{BA}(i\omega_n, \alpha) - i\mu' e^{-2\alpha}\delta_{AL}\delta_{BR} + i\mu' e^{2\alpha}\delta_{AR}\delta_{BL}}{(-i\omega_n - \kappa' - \Sigma_{LL})(-i\omega_n - \kappa' - \Sigma_{RR}) - (i\mu' e^{-2\alpha} - \Sigma_{LR})(-i\mu' e^{2\alpha} - \Sigma_{RL})}, \tag{30}$$

with Matsubara frequency $\omega_n = 2\pi(n+\frac{1}{2})/\beta$. The symmetry $G_{LL}(i\omega_n, \alpha) = G_{RR}(i\omega_n, \alpha)$, $G_{LR}(i\omega_n, \alpha) = -G_{RL}(i\omega_n, -\alpha)$, and $\Sigma_{LR}(i\omega_n, \alpha) = -\Sigma_{RL}(i\omega_n, -\alpha)$ are preserved under $T\bar{T}$ deformation.

Since the zero temperature theory produces no excitation on energy-momentum flow, a constant $a$ could be set as 0. The Hamiltonian term $H_0$ is negative in the thermal limit (6), and the square root has a real eigenvalue with $0 < \zeta < 1$. Note that the integral of function $\zeta$ is a constant in calculation, and the deformation parameter $\lambda$ is dimensionless. We can reduce these dimensionless parameters and seek how the model evolves with the new parameter $b$. Previous work has proved that the cSYK and MQ models have similar effects in thermal structure [44,47]. And we can focus more on how the thermodynamics change under deformation, instead of how the deformation depends on different parameters. In order to facilitate our calculation, we use the previous definition by mapping the parameter $b$ to the previous one. Symmetry between left and right models has been broken by non-Hermitian coupling [17]. As shown in Fig. 1, the decay rate of Green's function is shifted by $T\bar{T}$ deformation. In Fig. 1(a), we consider $T\bar{T}$ deformation on Hermitian Green's function; the maximal correlation becomes larger when we introduce $T\bar{T}$ deformation. And in Fig. 1(b), we involve the positive non-Hermitian parameter $\alpha$, which will strengthen the $G_{21}$ component while weaken $G_{12}$. According to $T\bar{T}$ deformation, the J and $\mu$ reparametrization converges to 0 when b is large enough, and the model should return to free fermions. Since non-Hermitian coupling breaks the symmetry between the left and right model, the off diagonal function $G_{21}$ will dominate when $\alpha \to +\infty$ [17]. And $T\bar{T}$ deformation will not break this behavior due to the relationship in (2) that is unbroken. Thermally, $T\bar{T}$ deformation provides the deformed Hamiltonian (6), indicating the projection of parameters $J, \mu, \kappa$ in (9), which builds additional correlation within and between left and right SYK model with nonlocal effects.

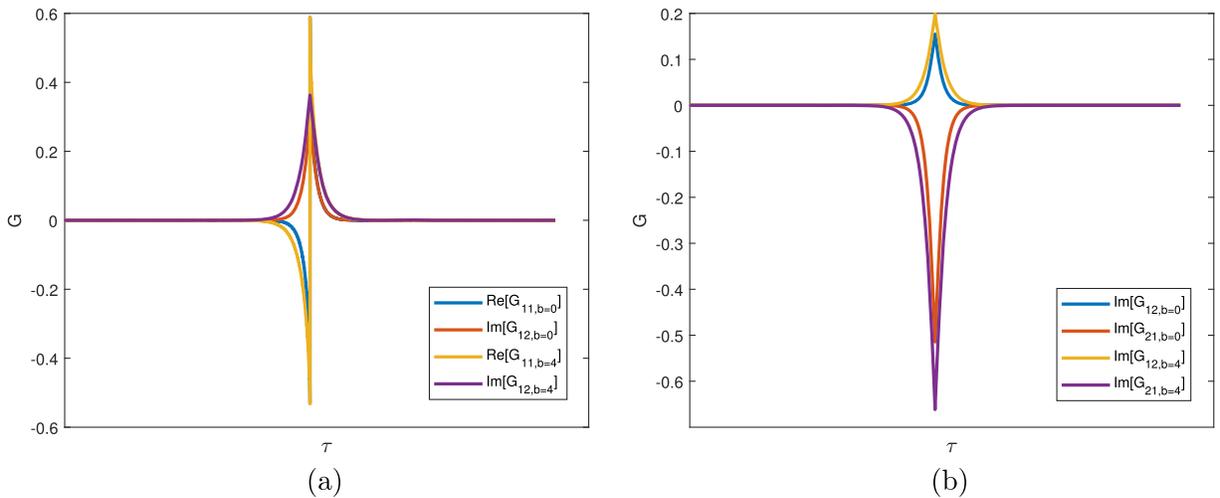

FIG. 1. (a) Undeformed $b = 0$ and deformed $b = 4$ Green's function in Hermitian case $\alpha = 0$ with fixed $J = 1$, $\mu = 0.3$, $T = 0.01$, $\kappa = 0$. (b) Undeformed $b = 0$ and deformed $b = 4$ Green's function in Hermitian case $\alpha = 0.3$ with fixed $J = 1$, $\mu = 0.3$, $T = 0.01$, $\kappa = 0$.

Substituting the saddle point solutions into the action, we obtain the free energy, as shown in Fig. 2,

$$\begin{aligned}\frac{F}{N} &= -T\frac{\log Z}{N} = T\frac{S_{\text{eff}}}{N} \\ &= -T\bigg[2\log 2 + \sum_{\omega_n}\log\frac{(-i\omega_n - \kappa' - \Sigma_{LL})(-i\omega_n - \kappa' - \Sigma_{RR}) - (i\mu' e^{-2\alpha} - \Sigma_{LR})(-i\mu' e^{2\alpha} - \Sigma_{RL})}{(i\omega_n)^2} \\ &\quad + \sum_{\omega_n}\bigg(\frac{3}{4}\Sigma_{LL}(i\omega_n,\alpha)G_{LL}(i\omega_n,\alpha) + \frac{3}{4}\Sigma_{RR}(i\omega_n,\alpha)G_{RR}(i\omega_n,\alpha) + \frac{3}{4}\Sigma_{LR}(i\omega_n,\alpha)G_{RL}(i\omega_n,\alpha) \\ &\quad + \frac{3}{4}\Sigma_{RL}(i\omega_n,\alpha)G_{LR}(i\omega_n,\alpha)\bigg)\bigg]. \end{aligned} \tag{31}$$

In order to plot the free energy, we decrease the temperature first, and then increase the temperature back to the high temperature. The free energy is not influenced by the non-Hermitian parameter $\alpha$ as shown in Fig. 2. The result is analogous to the undeformed case in [17]. After $T\bar{T}$ deformation, the free energy in the wormhole phase will increase, and the black hole entropy also increases due to the larger slope. And this is consistent with Nambu-Goto Hamiltonian and the irrelevant property of $T\bar{T}$ deformation. We can obtain a new phase-transition structure as the effective deformation parameter $b$ increases, and the coexistence region between wormhole phase and black hole phase is influenced by $b$ accordingly. Moreover, Fig. 2(b) shows the $T\bar{T}$ deformation makes the phase transition more distinguishable. The effective parameter $J$, $\mu$, and $\kappa$ in (9) decrease as $b$ increases, which is corresponding to a larger entropy and gapless two larger black holes, and the interaction $\mu$ for the wormhole phase should decrease. And we will discuss this holography picture in Sec. VI.

The phase diagram with dependence on $b$, $\mu$, and $T$ is numerically plotted in Fig. 3. At small $T$, the coupled system is holographically dual to an eternal traversable wormhole. At large $T$, the system reduces to the gapless two black hole phase.

The first-order Hawking-Page-like phase transition ends at a second-order critical point in Fig. 3(a). Our results show that $T\bar{T}$ deformation decreases the range of coexistence region between the wormhole phase and black hole phase due to a reparametrized $J$. And it shifts the transition critical points in both an Hermitian and non-Hermitian case from $(T_c = 0.269, \mu_c = 0.725, b = 0)$ to $(T_c = 0.246, \mu_c = 0.728, b = 0.1)$. As the deformation parameter $\lambda > 0$, the interaction $\mu$ also shifts to a small value, as shown in [17]. The thermodynamic transition temperatures gradually approach the wormhole saddle point as the effective deformation parameter $b$ increases, and the free energy saddle point of the black hole phase becomes larger than that in the undeformed theory. We have also plotted the phase diagram as a function of $b$ in Fig. 3(b), where the interaction $\mu$ is fixed. The coexistence region gradually shifts toward lower temperatures, and the transition temperature approaches the wormhole phase and then moves backward as $b$ increases. Additionally, the coexistence region is significantly enlarged due to the deformation parameter $b$. After a certain value of $b$, such as $b = 0.8$, $\mu = 0.4$, and $b = 1.4$, $\mu = 0.75$ in Fig. 3(b), the saddles of wormholes emerge with black holes, and the phase transition vanishes. However, the non-Hermitian parameter does not alter the $T\bar{T}$ deformation effects. We have proved that the wormhole-black hole thermodynamic structure is robust, not only in a Hermitian two-coupled SYK model, but also in a non-Hermitian two-coupled SYK model under the $T\bar{T}$ deformation. This result is predictable with a similar Hamiltonian Eq. (9) as reparametrization with $0 < \zeta < 1$.

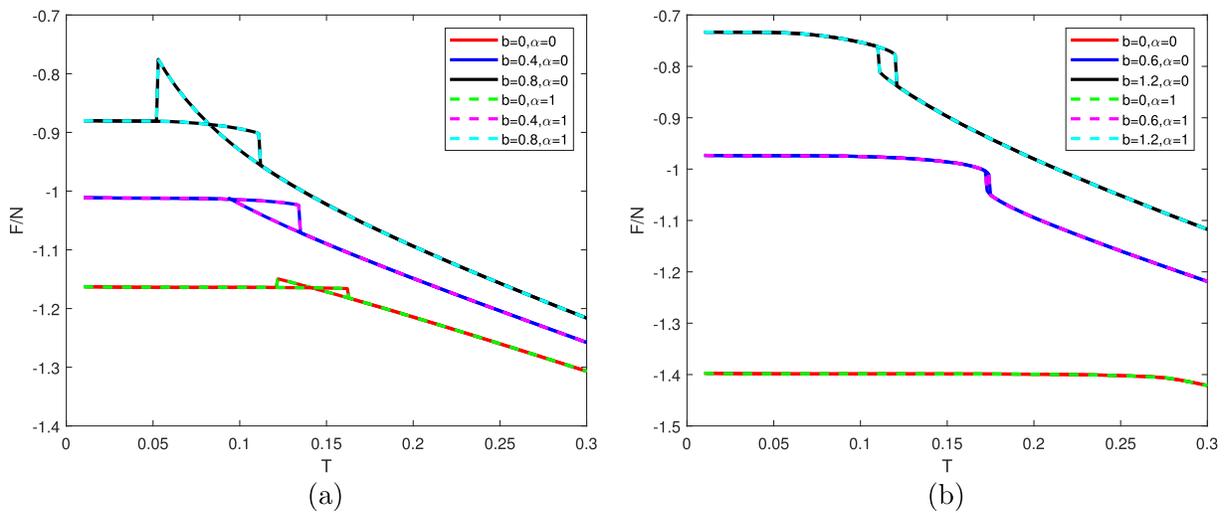

FIG. 2. The free energy as a function of $T$ in Hermitian case $\alpha = 0$ and non-Hermitian case $\alpha = 1$ for (a) different effective deformation parameter $b = 0, 0.4, 0.8$ with $J = 1, \mu = 0.4, \kappa = 0$. (b) different effective deformation parameter $b = 0, 0.6, 1.2$ with $J = 1, \mu = 0.75, \kappa = 0$.

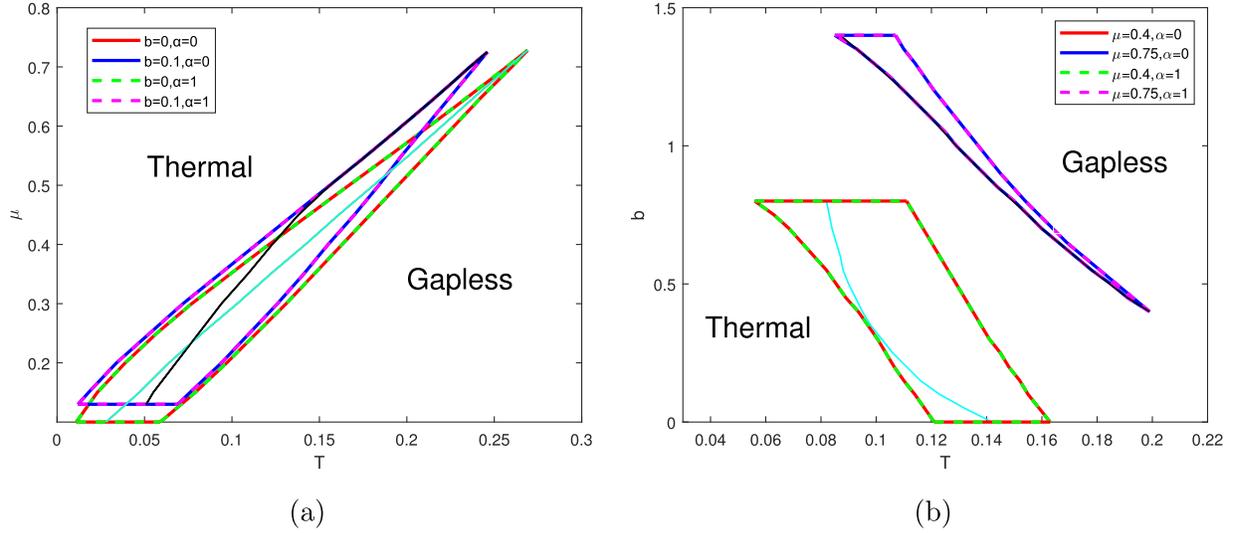

FIG. 3. Hermitian and non-Hermitian phase diagram with $J = 1$, $\kappa = 0$ (a) Undeformed $b = 0$ and $T\bar{T}$ deformation $b = 0.1$ on phase diagram. The coexist regions are surrounded; here the teal Green line is undeformed thermodynamic transition temperature, and the black line is deformed thermodynamic transition temperature. (b) Phase diagram as a function of the effective deformation parameter $b$ with couplings $\mu = 0.4$ and $\mu = 0.75$, the teal Green line and the black line are, respectively, $\mu = 0.4$ and $\mu = 0.75$.

In Eq. (28), the $T\bar{T}$ deformation is thermally equivalent to a reparametrization of parameters with $J^2/\mu$, $J^2/\kappa$ fixed, which appear in the effective action Eq. (9). And we provide several numerical examples in detail in Figs. 1–3. In order to propose an analytic proof for why the non-Hermitian parameter $\alpha$ does not affect the phase diagram, we consider a special transition of the Green's function in Eq. (10): $G'_{LL} \equiv G_{LL}$, $G'_{RR} \equiv G_{RR}$, $G'_{LR} \equiv \exp(-2\alpha)G_{LR}$, and $G'_{RL} \equiv \exp(2\alpha)G_{RL}$. Then, the effective action Eq. (27) with $G'_{AB}$ has the same form as that of a non-Hermitian theory,

$$\frac{S_E}{N} = -\log \det(\sigma_{AB} - \Sigma'_{AB}) - \int d\tau d\tau' \sum_{AB} [\Sigma'_{AB}(\tau,\tau')G'_{BA}(\tau',\tau) + 9J'^2 G'^2_{AB}(\tau,\tau')G'^2_{BA}(\tau',\tau)],$$

$$\Sigma'_{AB}(\tau,\tau') = -36 J'^2 G'^2_{AB}(\tau,\tau')G'_{BA}(\tau',\tau),$$

$$G'_{AB}(i\omega_n) = \frac{(-i\omega_n - \kappa')\delta_{AB} - \Sigma'_{BA}(i\omega_n,\alpha) - i\mu'\delta_{AL}\delta_{BR} + i\mu'\delta_{AR}\delta_{BL}}{(-i\omega_n - \kappa' - \Sigma'_{LL})(-i\omega_n - \kappa' - \Sigma'_{RR}) - (i\mu' - \Sigma'_{LR})(-i\mu' - \Sigma'_{RL})}. \quad (32)$$

The solution of this SD equation is independent of $\alpha$. So there is a statistical equivalence between the non-Hermitian and Hermitian systems, even though the non-Hermitian SYK model is not unitary.

## V. REVIVAL DYNAMIC

In this section, we investigate the dynamical properties and causality. Revival dynamic and transport properties could be obtained from Lorentzian-time Green's function in a Euclidean Keldysh time contour [7,48,49]. For this pseudo-Hermitian model, it is also convenient to preform analytically extended to the Lorentzian time via a Wick rotation and leads to a Wightman correlation function,

$$G^>_{AB}(t_1,t_2) = -i G_{AB}(it_1^-, it_2^+)$$
$$= -i \lim_{\epsilon \to +0} \langle \psi_A(\epsilon + it_1)\psi_B(-\epsilon + it_2)\rangle,$$
$$G^<_{AB}(t_1,t_2) = -G^>_{BA}(t_2,t_1). \quad (33)$$

Retarded and Advanced Green's function are also important in real time calculation,

$$G^R_{AB}(t_1, t_2) = \theta(t_1 - t_2)(G^>_{AB}(t_1, t_2) - G^<_{AB}(t_1, t_2)),$$
$$G^A_{AB}(t_1, t_2) = \theta(t_2 - t_1)(G^<_{AB}(t_1, t_2) - G^>_{AB}(t_1, t_2)), \quad (34)$$

where $\tau \to it$. The Schwinger-Dyson equation in Lorentzian time is constructed from the real time correlation function, and its retarded or advanced components is

$$i\partial_{t_1} G^>_{AB}(t_1, t_2) = \mu\epsilon_{AC} G^>_{CB}(t_1, t_2) + \int dt (\Sigma^R_{AC}(t_1, t) G^>_{CB}(t, t_2) + \Sigma^>_{AC}(t_1, t) G^A_{CB}(t, t_2)),$$
$$i\partial_{t_1} G^R_{AB}(t_1, t_2) - \mu\epsilon_{AC} G^R_{CB}(t_1, t_2) - \int dt (\Sigma^R_{AC}(t_1, t) G^R_{CB}(t, t_2) + \Sigma^A_{AC}(t_1, t) G^A_{CB}(t, t_2)) = \delta_{AB}\delta(t_1 - t_2), \quad (35)$$

with the self-energy term,

$$\Sigma^>_{AB}(t_1, t_2) = 36 J^2 G^>_{AB}(t_1, t_2) G^>_{AB}(t_1, t_2) G^>_{BA}(t_2, t_1),$$
$$\Sigma^R_{AB}(t_1, t_2) = \theta(t_1 - t_2)(\Sigma^>_{AB}(t_1, t_2) + \Sigma^>_{BA}(t_2, t_1)). \quad (36)$$

The Fourier transformation equation reduces to

$$G^R_{AB}(\omega) = \frac{(\omega - \kappa)\delta_{AB} - \Sigma^R_{BA}(\omega) - i\mu e^{-2\alpha}\delta_{aL}\delta_{bR} + i\mu e^{2\alpha}\delta_{aR}\delta_{bL}}{(\omega - \kappa - \Sigma^R_{LL})(\omega - \kappa - \Sigma^R_{RR}) - (i\mu e^{-2\alpha} - \Sigma^R_{LR})(-i\mu e^{2\alpha} - \Sigma^R_{RL})}. \quad (37)$$

In the Lorentz time form, $\omega_n \to i\omega$. The additional exponential divergence could be generated from Fourier transformation with real hyperbolic terms. We introduce the thermal decay rate, and its time strip by frequency partition function with a statistical temperature $T$ is

$$G^>_{AB}(\omega) = \frac{G^R_{AB}(\omega) - (G^R_{BA}(\omega))^*}{1 + \exp(-\beta\omega)}. \quad (38)$$

In an Hermitian system, a Lorentz-time Green's function could be reduced to pure real or pure imaginary,

$$G^>_{LL}(\omega) = \frac{2i\text{Im}(G^R_{LL}(\omega))}{1 + \exp(-\beta\omega)},$$
$$G^>_{LR}(\omega) = \frac{2\text{Re}(G^R_{LR}(\omega))}{1 + \exp(-\beta\omega)}. \quad (39)$$

The non-Hermitian results are similar; we can also utilize the process in Eq. (32). The non-Hermitian parameter $\alpha$ will generate additional coefficients in Eq. (38) without changing the equation.

$T\bar{T}$ deformation in Lorentz time could be performed exactly in the same way in an Euclidean situation, except for switching the original Euclidean representation to the Lorentzian one. From the $T\bar{T}$ flow in Lorentz-time action with corresponding Schwinger-Dyson (SD) equations and correlation function, the auxiliary field we have introduced before in Euclidean space time turns out to be

$$i\frac{S_{\text{total}}}{N} = i\frac{S_L + S_R + S_{\text{int}}}{N}$$
$$= \log \det(\omega\delta_{AB} - \Sigma_{AB}(\omega)) - \frac{1}{2}\int d\tau d\tau' \sum_{AB}[\Sigma_{AB}(\tau, \tau')G_{BA}(\tau', \tau) - 9J'^2 G^2_{AB}(\tau, \tau')G^2_{BA}(\tau', \tau)]$$
$$+ \frac{i\mu'}{2}\int d\tau d\tau'[G_{LR}(\tau, \tau')\delta(\tau - \tau') - G_{RL}(\tau', \tau)\delta(\tau' - \tau)]$$
$$- \frac{\kappa'}{2}\int d\tau d\tau'[G_{LL}(\tau, \tau')\delta(\tau - \tau') + G_{RR}(\tau', \tau)\delta(\tau' - \tau)]. \quad (40)$$

In order to reduce additional elements with an Euclidean symbol, it is natural to transform the function along another Keldysh contour into a Lorentzian,

$$\Sigma = \Sigma^{>} - \Sigma^{R}, \qquad G = G^{A} + G^{>}. \tag{41}$$

The auxiliary parameter $\zeta$ should be switched to a real time patch through the real time SD equation,

$$\zeta^{-2} = 1 + b \int dt [\sum_{AB} G_{AB}^{2}(t) G_{BA}^{2}(-t) + \mu(-e^{2\alpha} G_{LR}(t) \delta(-t) + e^{-2\alpha} G_{RL}(t) \delta(-t)) + i\kappa(G_{LL}(t) \delta(-t) + G_{RR}(t) \delta(-t))] \tag{42}$$

After involving the contoured function, $\zeta$s are now written in the real time correlation and its advanced component,

$$\zeta^{-2} = 1 + b \int dt \bigg( \sum_{AB} (G_{AB}^{A}(t) + G_{AB}^{>}(t))^{2} (G_{BA}^{A}(-t) + G_{BA}^{>}(-t))^{2}$$
$$+ \mu(-e^{2\alpha} (G_{LR}^{A}(t) + G_{LR}^{>}(t)) \delta(-t) + e^{-2\alpha} (G_{RL}^{A}(t) + G_{RL}^{>}(t)) \delta(-t))$$
$$- i\kappa((G_{LL}^{A}(t) + G_{LL}^{>}(t)) \delta(-t) + (G_{RR}^{A}(t) + G_{RR}^{>}(t)) \delta(-t)) \bigg). \tag{43}$$

According to transmission symmetry between retarded and advanced correlation, we simply shift the expression into the retarded form,

$$\zeta^{-2} = 1 + b \int dt \bigg( \sum_{AB} (-G_{AB}^{R}(-t) + G_{AB}^{>}(t))^{2} (-G_{BA}^{R}(t) + G_{BA}^{>}(-t))^{2}$$
$$+ \mu(-e^{2\alpha} (-G_{LR}^{R}(-t) + G_{LR}^{>}(t)) \delta(-t) + e^{-2\alpha} (-G_{RL}^{R}(-t) + G_{RL}^{>}(t)) \delta(-t))$$
$$- i\kappa((-G_{LL}^{R}(-t) + G_{LL}^{>}(t)) \delta(-t) + (-G_{RR}^{R}(-t) + G_{RR}^{>}(t)) \delta(-t)) \bigg). \tag{44}$$

We used to utilize Fourier transformation to detour an infinite proper time integral. However, Matsubara Fermion frequencies in an usual many-body quantum system recover a divergent result in Lorentz-time equations. For a numerical need, a real-time cutoff could be performed without violating the limited causality in both frequency space and Lorentz time space. Performing a longtime cutoff would be a good attempt for the numerical approximation method [50]. The discrete numerical timescales are now given by

$$t_{m} = m t_{\max}, \qquad m = (-\Lambda, -\Lambda + 1, \ldots, \Lambda - 1)/\Lambda. \tag{45}$$

The conjugated discrete frequency is

$$\omega_{n} = \pi n / t_{\max}, \qquad n = (-\Lambda, -\Lambda + 1, \ldots, \Lambda - 1). \tag{46}$$

It is natural to assume that the real-time correlation function should decay to 0 in a long time limit, and the sector $t > t_{\max}$ has been dropped. We have ignored the IR and UV behaviors. After discrete Fourier transformation, we obtain $\zeta$ with the frequency,

$$\zeta^{-2} = 1 + b/(2t_{\max}) \sum_{\omega_{n}} \bigg[ \sum_{AB} (-G_{AB}^{R}(-\omega_{n}) + G_{AB}^{>}(\omega_{n}))^{2} (-G_{BA}^{R}(-\omega_{n}) + G_{BA}^{>}(\omega_{n}))^{2}$$
$$+ \mu(-e^{2\alpha} (-G_{LR}^{R}(\omega_{n}) + G_{LR}^{>}(\omega_{n})) + e^{-2\alpha} (-G_{RL}^{R}(\omega_{n}) + G_{RL}^{>}(\omega_{n})))$$
$$- i\kappa((-G_{LL}^{R}(\omega_{n}) + G_{LL}^{>}(\omega_{n})) + (-G_{RR}^{R}(\omega_{n}) + G_{RR}^{>}(\omega_{n}))) \bigg]. \tag{47}$$

Terms with $\omega \gg 1/t_{\max}$ are insignificant in the equation. Notice that the temperature $T$ dependent $\zeta$ in Euclidean time is somehow replaced by inverse maximal Lorentzian cutoff $1/2t_{\max}$. The temperature now appears in statistical real time decay spectral. However, in this approximate theory, the numerical relation $1/2t_{\max}$ is not arbitrary, since the time cutoff $1/2t_{\max}$ and inverse temperature $\beta$ should match each other. We numerically calculate the real time propagator of the transition point in Fig. 4.

And it is significant to choose the critical transition point between wormhole phase and two-black hole phase. More explicitly, we choose the maximal transmission amplitude $|2G_{LR}| \approx 1$. The $T\bar{T}$ deformation shifts the thermal phase structure as we calculated in the previous section. The result of the wormholelike solution at the transition point makes the oscillation vanish and gradually transforms into a black hole after utilizing the $T\bar{T}$ deformation. We have also found a solution that the wormhole revival function with $\mu = 0.3$ greatly enhances the $G_{LR}$ component and replaces the role of $G_{LL}$ does when $\mu = 0.1$ or smaller. We shall focus on the dynamical decays for interaction parts. In Figs. 4(b) and 4(d), the deformation will shift the phase diagram from original position, the temperature here is no longer at transition point, and the density is not convergent. In the real time non-Hermitian case, the symmetry between the propagator of L-R and R-L is broken. Here, we have plotted a observably example at the critical point In Figs. 4(c) and 4(d), the weak $T\bar{T}$ deformation $b = 0.001$ will observably decrease the oscillation. And this result agrees with phase structure in

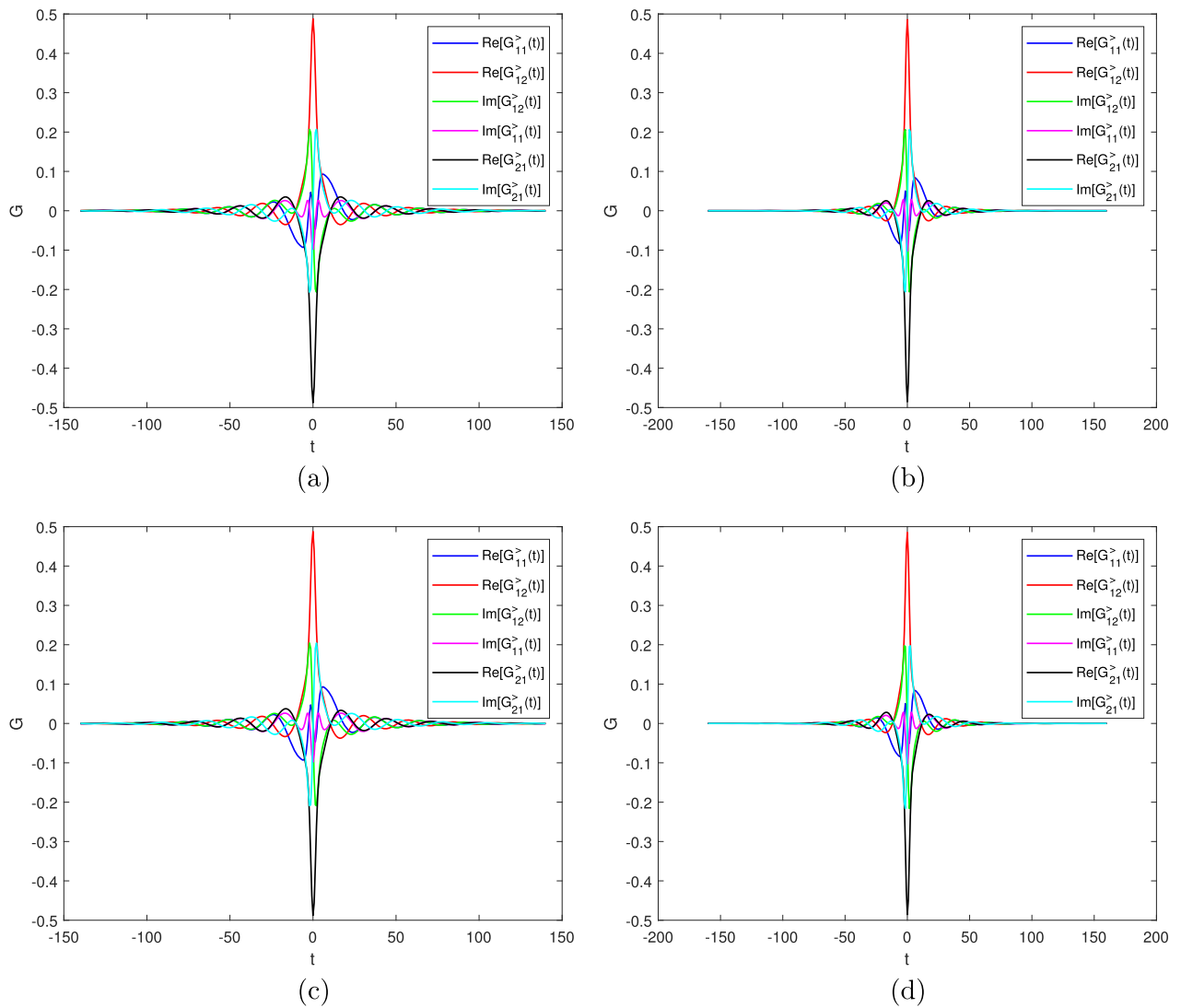

FIG. 4. Revival dynamic in wormholelike critical point with $J = 1$, $\mu = 0.3$, $T = 0.131$, $\lambda = 200000$ (a) Real time propagator with Hermitian $\alpha = 0$, $t_{\max} = 140000$, $b = 0$. (b) Real time propagator with Hermitian $\alpha = 0$, $t_{\max} = 160000$, $b = 0.001$. (c) Real time propagator with non-Hermitian $\alpha = 0.01$, $t_{\max} = 140000$, $b = 0$. (d) Real time propagator with non-Hermitian $\alpha = 0.01$, $t_{\max} = 160000$, $b = 0.001$.

Sec. IV. When we further increase the parameter $b$, the oscillation decreases and turns into the black hole solution, and we can also find a correspondence in Fig. 3. We have also introduce a non-Hermitian parameter $\alpha = 0.01$ as a example, which enhances the real component of $G_{RL}$ in advanced time $t < 0$ and weakens it in retarded time $t > 0$, while the $G_{LR}$ component is completely opposite. In black hole phase, the numerical limit $t_{\max}$ increases, and it is numerically equal to decreasing the temperature to a wormhole oscillation solution. According to the results in Sec. IV, the effective deformation parameter $b$ will change the phase diagram. Here, the corresponding revival dynamics of phases at the transition point can be transformed into each other due to the $T\bar{T}$ deformation.

## VI. DEFORMED $N$AdS$_2$ SPACETIME AND HOLOGRAPHY

In this section, we will investigate the $T\bar{T}$ deformation on $N$AdS$_2$ spacetime. It is natural to consider a 2D gravitational $T\bar{T}$ deformation in bulk theory [23]. However, it seems more convenient to process with a 1D quadratic stress tensor.

We suppose $N$AdS$_2$ geometry with Einstein gravity, which can be described as Jackiw-Teitelboim(JT) gravity [51,52],

$$S = \frac{\phi_0'}{2}\left[\int \sqrt{g}\mathcal{R} + 2\int_{\text{Bdy}}\sqrt{g^0}\mathcal{K}\right] + \frac{1}{2}\left[\int \phi'\sqrt{g}(\mathcal{R}+2) + 2\int_{\text{Bdy}}\phi_b'\sqrt{g^0}\mathcal{K}\right] + S_{\text{matter}}$$
$$= \frac{1}{2}\left[\int \sqrt{g}(\phi_0\mathcal{R} + \phi) + 2\int_{\text{Bdy}}\sqrt{g^0}\phi_b\mathcal{K}\right] + S_{\text{matter}}, \tag{48}$$

with a constraint on boundary dilatons [50,53] $\phi'_{\text{Bdy}} = \phi'_b$. The first and second parts in (48) denote gravitational action on NAdS spacetime with curvatures. The third term contains a possible additional matter field and the interaction term in coupled SYK model. $T\bar{T}$ deformation in AdS has been well proved as a boundary cutoff in [54], and we can follow the technique [23] and expand it into NAdS with additional boundary dilatons. Holographic $T\bar{T}$ deformation in AdS spacetime has been proved to be a rigid radial cutoff with finite surface. However, this physical interpretation does not coexist with any additional matter field, because the cutoff metric does not contain a $T\bar{T}$ flow equation on the sphere. For simplicity, we have introduced a modification into dilaton terms with diagonal gauge metric,

$$ds^2 = N^2(r)d\tau^2 + \frac{dr^2}{N^2(r)}. \tag{49}$$

Here, AdS metric element $N^2$ could be written as

$$N^2(r) = f(\phi_0)\left(1 - \frac{2M}{\phi_r(r)f(\phi_0)}\right), \tag{50}$$

$$\phi_0(r) = r\phi_r(r). \tag{51}$$

$f(\phi_0)$ depends on the variable coordinate $\phi_r$ and bulk integral of $\phi$, and $M$ is a constant with dimension of mass. Function $f$ is written as

$$f(\phi_0) = \frac{2}{\phi_r^2}\int_{\phi_0}\phi(x)dx. \tag{52}$$

If we consider a finite Dirichlet radial cutoff, the quasilocal bulk energy solution is given by

$$E_{\text{bulk}} = \phi_r r_c\left(1 - \frac{1}{r_c}\sqrt{f(\phi_0(r_c)) - 2M/\phi_r}\right), \tag{53}$$

which is familiar to Nambu-Goto-like Hamiltonian. We may expect a $T\bar{T}$ deformation formalism. We now seek a method to perform the $T\bar{T}$ deformation equation with

$$\frac{\partial S}{\partial \lambda} = \int d\tau \frac{T^2}{1/2 - 2\lambda T}. \tag{54}$$

In order to find the $T\bar{T}$ operator with energy-momentum flow, we can define a Brown-York stress scalar with renormalization. Two-dimensional $T\bar{T}$ flow is obtained by variation of the metric $g^0$ and bulk EFT dilaton $\phi_0$. And the metric and stress tensor are passively transformed. The boundary interaction term is added to neutral Maldacena and Qi's (MQ) model [6] with a non-Hermitian interaction [17],

$$S_{\text{int}} = g\sum_i \int du(e^{-2\alpha}O_L^{i\dagger}(u)O_R^i(u) - e^{2\alpha}O_R^{i\dagger}(u)O_L^i(u)). \tag{55}$$

The semiclassical interaction operators $O$ are set with conformal dimension of $\Delta$. The constant $g$ marks the stress with conformal limit. We define

$$g_{\tau\tau}^0 = r_c^2 \gamma_{\tau\tau},$$

$$T^{\tau\tau} = r_c \tilde{T}^{\tau\tau} = \frac{2}{\sqrt{g^0}} \frac{\delta S_{\text{total}}}{\delta g^0} = \frac{2}{\sqrt{g^0}} \frac{\delta(2S + S_{\text{int}})}{\delta g^0},$$

$$T^\phi = r_c^{-2} \tilde{T}^\phi = \frac{1}{\sqrt{g^0}} \frac{\delta S_{\text{total}}}{\delta \phi_0} = \frac{1}{\sqrt{g^0}} \frac{\delta(2S + S_{\text{int}})}{\delta \phi_0}. \quad (56)$$

Additional matter field in (45) would break the symmetry of flow equation. However, the interaction term in a coupled SYK model [6] only includes boundary time dependence. And in the finite radius of bulks, we have

$$\frac{\partial S_{\text{matter}}}{\partial T^{\mu\nu}} = \frac{\partial S_{\text{int}}}{\partial T^{\mu\nu}} = 0, \quad (57)$$

which means the $T\bar{T}$ deformation on $NAdS_2$ with boundary interaction will be a global effect at finite coordinates $\mu$ and $\nu$. And this condition does not contradict the $T\bar{T}$ equation (6). We can involve the boundary conditions of $NAdS_2$ into tensors without changing the equations. According to the definition of $T\bar{T}$ deformation, the $NAdS_2$ flow equation could be performed with a similar $T\bar{T}$ operator.

The flow equation of effective action with on shell Hamiltonian constraint can be transformed into pure JT gravity globally at finite, and the boundary interaction under $T\bar{T}$ deformation will not cause a nontrivial local effect in spacetime. And we have

$$r_c \frac{\partial}{\partial r_c} S_{\text{EFT}} = \int d\tau \sqrt{\gamma} \left( \frac{-\tilde{T}_\tau^\tau + \phi_r^2 r_c^4 + \phi_r r_c^3 \phi(r_c \phi_r)}{\phi_r r_c^2 - \tilde{T}_\tau^\tau} \right). \quad (58)$$

Notice that the flow equation now exactly has the same mathematical form as $T\bar{T}$ deformation. By comparing formulas we have mentioned, we obtain the deformation parameter $\lambda$ in $NAdS_2$ spacetime,

$$\lambda = \frac{1}{4\phi_r r_c^2}\Big|_{r_c=\text{finite}}. \quad (59)$$

We are more concerned with bulks, and it is natural to remove the neighborhood of the boundaries. Thus we can write the flow equation with boundary cutoff in the $T\bar{T}$ deformation form,

$$\frac{\partial S_{\text{EFT}}}{\partial \lambda} = \int d\tau \sqrt{\gamma} \left( \frac{T - (1 + (r_c\phi_r)^{-1}\phi(r_c\phi_r))/(16\lambda^2)}{1/2 - 2\lambda T} \right). \quad (60)$$

Here, we have set the diagonal $T\bar{T}$ stress tensor with

$$T = \tilde{T}_\tau^\tau. \quad (61)$$

The deformation equation could also be written with the Hamiltonian,

$$\frac{\partial E}{\partial \lambda} = \frac{E^2 - \left(1 + \frac{\sqrt{4\phi_r \lambda}}{\phi_r}\phi\left(\frac{\phi_r}{\sqrt{4\phi_r \lambda}}\right)\right)/(16\lambda^2)}{1/2 - 2\lambda E}, \quad (62)$$

which may generates a simple Nambu-Goto solution,

$$E(\lambda) = \frac{1}{4\lambda}\left(1 - \sqrt{4\phi_r \lambda} f\left(\frac{\phi_r}{\sqrt{4\phi_r \lambda}}\right) - 8\lambda E_0\right). \quad (63)$$

And when we focus on the cutoff space, additional parts outside the cutoff sphere will be indistinguishable from NAdS boundary term. So we have proved that the $T\bar{T}$ deformation of $NAdS_2$ bulk metric with boundary interaction is equal to a finite radial Dirichlet boundary cutoff. However, this does not mean to abandon the geometry between the deformed radial boundary and the original $NAdS_2$ boundary which becomes indistinguishable. We focus on the dynamics and energy-momentum flow on the radial Dirichlet sphere, which is exactly equal to the $T\bar{T}$ deformation. We may find a similar flow equation as $T\bar{T}$ deformation in NAdS spacetime as dual coupled SYK does. Since the interaction term is irrelevant to a metric tensor $g$, this additional term does not break the physical interpretation of $T\bar{T}$ deformation. One notable thing is that the interaction Hamiltonian $H_{\text{int}}$ definitely plays a role in the $T\bar{T}$ flow, but globally. The interaction has already modified the natural boundary, and the $T\bar{T}$ deformation on these modified boundaries are still equal to a radial cutoff.

The bulk theory reduces to Schwarzian form by a time reparametrization and leads to symmetry breaking. We would like to involve the $T\bar{T}$ deformation to effective Schwarzian theory without breaking holographic duality. We can introduce

$$S_E = -C \int d\tau (\text{Sch}(h,\tau) + W(h')), \quad (64)$$

where $C$ is a physical coefficient. And it can be defined as a dilaton reparametrization on the boundary $\phi'_{\text{Bdy}} = \phi'_b = \frac{\bar{\phi}_r}{\epsilon}$ and $ds|_{\text{Bdy}} = \frac{d\tau}{\epsilon}$ in (48). Taking $\epsilon$ to be infinitesimal, we have $C = \bar{\phi}_r$. In the dual SYK effective theory, $C$ could also be some quantum constant which depends on the four point correlation function $C = N\alpha_S$. In order to fix the low energy effective action in [6], we introduce an arbitrary $W(h')$ function of $h'$ as a second term. Taking $W(h')$ to zero, the theory reduces to a pure Schwarzian derivative,

$$\text{Sch}(h,\tau) = \left(\frac{h''}{h'}\right)' - \frac{1}{2}\left(\frac{h''}{h'}\right)^2. \quad (65)$$

One can consider the time reparametrization of a radial cutoff metric, but it seems more convenient to deform Schwarzian action directly. We can also extend the technique of $T\bar{T}$ deformed Schwarzian action in [23] to effective action of the MQ model. The first step toward $T\bar{T}$ deformation is to find the energy-momentum tensor and establish its flow equation, and we can pick a canonical coordinates with Ostrogradsky formalism [55],

$$q_1 = h, \qquad q_2 = h' = e^{\varphi}, \tag{66}$$

with canonical momentums,

$$p_1 = \frac{\partial L}{\partial h'} - \frac{d}{d\tau}\left(\frac{\partial L}{\partial h''}\right) = C\left(\frac{h''^2}{h'^3} - \frac{h'''}{h'^2} - w(h')\right),$$
$$p_2 = \frac{\partial L}{\partial h''} = C\frac{h''}{h'^2}, \tag{67}$$

with $w(h') = \partial W(h')/\partial h'$. The Hamiltonian is invariant under coordinate transformation,

$$H = \frac{p_2^2 q_2^2}{2C} + \frac{C}{2}q_2^2 + p_1 q_2. \tag{68}$$

Note that the $T\bar{T}$ deformation could be projected via Hamiltonian form,

$$H(\lambda) = \frac{1}{4\lambda}(1 - \sqrt{1 - 8\lambda H_0}). \tag{69}$$

We obtain the deformed Lagrangian by Legendre transformation with coordinates $\tau \to -i\tau, q_2 \to iq_2$,

$$L_E(\lambda) = -\frac{(h' - e^{\varphi})}{8\lambda h' e^{\varphi}} + \frac{C}{2}\frac{e^{\varphi}}{h'}(\varphi'^2 - h'f(h')). \tag{70}$$

First, we perform a deformation to a low energy effective SYK theory, which has an SL(2) boundary reparametrization symmetry and a pure Schwarzian derivative without an additional function $W(h')$,

$$S = -C\int du\{t_P(u), u\}. \tag{71}$$

$t_P$ is a possible reparametrization of $u$. And we can apply a simple Lorentzian solution,

$$S_E = -C\int d\tau\left\{\tanh\frac{h(\tau)}{2}, \tau\right\}. \tag{72}$$

Corresponding deformed Lagrangian are parametrized from a boundary,

$$L_E(\lambda) = -\frac{\left(\tanh\frac{h(\tau)'}{2} - e^{\varphi}\right)^2}{8\lambda \tanh\frac{h(\tau)'}{2}e^{\varphi}} - \frac{C}{2}\frac{e^{\varphi}}{\tanh\frac{h(\tau)'}{2}}(\varphi'^2). \tag{73}$$

After involving constant $C$, we have the holographic effective action,

$$L_E(\lambda) = -\frac{\left(\tanh\frac{h(\tau)'}{2} - e^{\varphi}\right)^2}{8\lambda \tanh\frac{h(\tau)'}{2}e^{\varphi}} - \frac{\bar{\phi}_r}{2}\frac{e^{\varphi}}{\tanh\frac{h(\tau)'}{2}}(\varphi'^2), \tag{74}$$

which can be written as

$$L_E(\lambda) = -\frac{\left(\tanh\frac{h(\tau)'}{2} - e^{\varphi}\right)^2}{8\lambda \tanh\frac{h(\tau)'}{2}e^{\varphi}} - \frac{N\alpha_S}{2}\frac{e^{\varphi}}{\tanh\frac{h(\tau)'}{2}}(\varphi'^2). \tag{75}$$

The conformal transformation with additional complex U(1) symmetry is given [56,57]

$$\tilde{G}_{AB}(\tau, \tau\prime) = [h'_A(\tau)h'_B(\tau')]^{\Delta}G_{AB}(h_A(\tau), h_B(\tau'))$$
$$\times e^{i(\varphi_A(\tau) - \varphi_B(\tau'))},$$
$$\tilde{\Sigma}_{AB}(\tau, \tau\prime) = [h'_A(\tau)h'_B(\tau')]^{1-\Delta}G_{AB}(h_A(\tau), h_B(\tau'))$$
$$\times e^{i(\varphi_A(\tau) - \varphi_B(\tau'))}. \tag{76}$$

Left and right Schwarzian actions have an U(1) anomalous term to preserve transformation symmetry,

$$S = -N\alpha_S \int d\tau\left\{\tanh\frac{h(\tau)}{2}, \tau\right\}$$
$$+ \frac{NK}{2}\int d\tau(\varphi'(\tau) + i\varepsilon h'(\tau))^2, \tag{77}$$

where the parameter $\varepsilon$ is generated by a U(1) charge with a generator $\varphi$, and $\alpha_S$ is determined by a four-point and higher order calculations. $K$ appears to be coefficient of compressibility. The general reparametrized Schwarzian action nearly performs $SL(2, \mathbb{R}) \times U(1)$ Virasoro-Kac-Moody symmetry,

$$\delta h_L = \epsilon^0 + \epsilon^+ e^{ih_L} + \epsilon^- e^{-ih_L},$$
$$\delta h_R = \epsilon^0 - \epsilon^+ e^{ih_R} - \epsilon^- e^{-ih_R},$$
$$\delta\varphi = -i\varepsilon\delta h + \epsilon, \tag{78}$$

with infinitesimal displacements $\epsilon^0$, $\epsilon^+$, and $\epsilon^-$. And the corresponding function $w(h')$ in $SL(2, \mathbb{R}) \times U(1)$ Schwarzian theory,

$$w(h') = \frac{1}{C}\frac{\partial W(h')}{\partial h'} = \frac{i\varepsilon NK}{C}(\varphi'(\tau) + i\varepsilon h'(\tau)). \tag{79}$$

This result leads to the deformed Lagrangian,

$$L_{E,A}(\lambda) = -\frac{\left(\tanh\frac{h_A(\tau)'}{2} - e^{\varphi}\right)^2}{8\lambda \tanh\frac{h_A(\tau)'}{2} e^{\varphi}} - \frac{N\alpha_S}{2} \frac{e^{\varphi}}{\tanh\frac{h_A(\tau)'}{2}} \left(\varphi'^2 - \tanh\frac{h_A(\tau)'}{2} \left(\frac{i\varepsilon K}{\alpha_S}(\varphi'(\tau) + i\varepsilon h'_A(\tau))\right)\right), \tag{80}$$

and the coupling part of effective action of two SYK models is given as

$$S_{\text{int}} = \frac{\mu}{2} \int d\tau \left[\frac{B h'_L(\tau) h'_R(\tau)}{\cosh^2\left(\frac{h_L(\tau) - h_R(\tau)}{2}\right)}\right]^{\Delta}. \tag{81}$$

$B$ denotes a reparametrization in a low energy correlation function with [6]. Since we $T\bar{T}$ deform the total action with decoupled models and their interactions, these are described by Schwarzian derivative and reparametrized interaction operators,

$$S_{\text{total}} = 2S_A + S_{\text{int}}. \tag{82}$$

Parameter A refers to L and R models. The corresponding function $w$ becomes

$$w_A(h') = \frac{i\varepsilon K}{\alpha_S}(\varphi'(\tau) + i\varepsilon h'_A(\tau)) + \frac{\Delta B^{\Delta} \mu}{2N\alpha_S \cosh^{2\Delta}\left(\frac{h_L(\tau) - h_R(\tau)}{2}\right)} [h'_L(\tau) h'_R(\tau)]^{\Delta - \frac{1}{2}}, \tag{83}$$

and we obtain the corresponding Lagrangian,

$$L_{E,A}(\lambda) = -\frac{\left(\tanh'\frac{h_A(\tau)}{2} - e^{\varphi}\right)^2}{8\lambda \tanh'\frac{h_A(\tau)}{2} e^{\varphi}} - N\alpha_S \frac{e^{\varphi}}{\tanh'\frac{h_A(\tau)}{2}}$$
$$\times \left(\varphi'^2 - \tanh'\frac{h_A(\tau)}{2} \left(\frac{i\varepsilon K}{\alpha_S}(\varphi'(\tau) + i\varepsilon h'_A(\tau)) + \frac{\Delta B^{\Delta} \mu}{2N\alpha_S \cosh^{2\Delta}\left(\frac{h_L(\tau) - h_R(\tau)}{2}\right)} [h'_L(\tau) h'_R(\tau)]^{\Delta - \frac{1}{2}}\right)\right). \tag{84}$$

We can also incorporate the additional U(1) parameter $\varphi$ and non-Hermitian parameter $\alpha$ into the interaction term,

$$S_{\text{int}} = \frac{\mu}{2} \int d\tau \left[\frac{B h'_L(\tau) h'_R(\tau)}{\cosh^2\left(\frac{h_L(\tau) - h_R(\tau)}{2}\right)}\right]^{\Delta} \cosh\left(\varepsilon h_L(\tau) - \varepsilon h_R(\tau)\right)[\exp\left(i(\varphi_L - \varphi_R) - 2\alpha\right) + \exp\left(-i(\varphi_L - \varphi_R) + 2\alpha\right)]. \tag{85}$$

For function $w$, this variation is equal to the addition of an external structure,

$$w_A(h') = \frac{i\varepsilon K}{\alpha_S}(\varphi'(\tau) + i\varepsilon h h'_A(\tau)) + \frac{\Delta B^{\Delta} \mu}{2N\alpha_S \cosh^{2\Delta}\left(\frac{h_L(\tau) - h_R(\tau)}{2}\right)} [h'_L(\tau) h'_R(\tau)]^{\Delta - \frac{1}{2}}$$
$$\times (\cosh\left(\varepsilon h_L(\tau) - \varepsilon h_R(\tau)\right) \times [\exp\left(i(\varphi_L - \varphi_R) - 2\alpha\right) + \exp\left(-i(\varphi_L - \varphi_R) + 2\alpha\right)]), \tag{86}$$

the corresponding Lagrangian also includes the external exponent,

$$L_{E,A}(\lambda) = -\frac{\left(\tanh'\frac{h_A(\tau)}{2} - e^{\varphi}\right)^2}{8\lambda \tanh'\frac{h_A(\tau)}{2} e^{\varphi}} - N\alpha_S \frac{e^{\varphi}}{\tanh'\frac{h_A(\tau)}{2}}$$
$$\times \left(\varphi'^2 - \tanh'\frac{h_A(\tau)}{2} \left(\frac{i\varepsilon K}{\alpha_S}(\varphi'(\tau) + i\varepsilon h'_A(\tau)) + \frac{\Delta B^{\Delta} \mu}{2N\alpha_S \cosh^{2\Delta}\left(\frac{h_L(\tau) - h_R(\tau)}{2}\right)} [h'_L(\tau) h'_R(\tau)]^{\Delta - \frac{1}{2}}\right)\right.$$
$$\times \left.(\cosh\left(\varepsilon h_L(\tau) - \varepsilon h_R(\tau)\right) \times [\exp\left(i(\varphi_L - \varphi_R) - 2\alpha\right) + \exp\left(-i(\varphi_L - \varphi_R) + 2\alpha\right)])\right). \tag{87}$$

Now we switch to gravitational Schwarzian action with non-Hermitian boundary interaction, which leads to a renormalized effective action,

$$S_{\text{int}} = \frac{gN}{2^{2\Delta}} \int d\tau \left[\frac{h'_L(\tau)h'_R(\tau)}{\cosh^2\left(\frac{h_L(\tau)-h_R(\tau)}{2}\right)}\right]^\Delta \cosh\left(\varepsilon h_L(\tau) - \varepsilon h_R(\tau)\right)[\exp\left(i(\varphi_L - \varphi_R) - 2\alpha\right) + \exp\left(-i(\varphi_L - \varphi_R) + 2\alpha\right)]. \quad (88)$$

And gravitational Lagrangian is given as

$$\begin{aligned}
L_{E,A}(\lambda) = &-\frac{\left(\tanh'\frac{h_A(\tau)}{2} - e^\varphi\right)^2}{8\lambda\tanh'\frac{h_A(\tau)}{2}e^\varphi} - \bar{\phi}_r \frac{e^\varphi}{\tanh'\frac{h_A(\tau)}{2}} \\
&\times \left(\varphi'^2 - \tanh'\frac{h_A(\tau)}{2}\left(\frac{i\varepsilon\tilde{K}}{\bar{\phi}_r}(\varphi'(\tau) + i\varepsilon h'_A(\tau)) + \frac{gN\Delta}{2^{2\Delta}\bar{\phi}_r\cosh^{2\Delta}\left(\frac{h_L(\tau)-h_R(\tau)}{2}\right)}[h'_L(\tau)h'_R(\tau)]^{\Delta-\frac{1}{2}}\right) \right. \\
&\left. \times \left(\cosh\left(\varepsilon h_L(\tau) - \varepsilon h_R(\tau)\right)\right) \times [\exp\left(i(\varphi_L - \varphi_R) - 2\alpha\right) + \exp\left(-i(\varphi_L - \varphi_R) + 2\alpha\right)]\right).
\end{aligned} \quad (89)$$

The deformation is relevant to mathematical formalisms $h$ and $h'$, but the $T\bar{T}$ deformation does not affect its physical coefficients C; $T\bar{T}$ deformation on effective action would not break holography duality,

$$\bar{\phi}_r = N\alpha_S, \qquad \frac{\mu B^\Delta}{2} = \frac{gN}{2^{2\Delta}}. \quad (90)$$

In $T\bar{T}$ deformed Lagrangian (87), the SL(2) symmetry has been broken. However, the reparametrization symmetric U(1) relation between $h'$ and $\varphi$ is preserved.

We can also simplify the total action with symmetry $h_L = h_R = h(\tau)$ and $\varphi_L = \varphi_R - 2i\alpha = \varphi(\tau)$,

$$\begin{aligned}
L_E(\lambda) = &-\frac{\left(\tanh'\frac{h(\tau)}{2} - e^\varphi\right)^2}{8\lambda\tanh'\frac{h(\tau)}{2}e^\varphi} - N\alpha_S \frac{e^\varphi}{\tanh'\frac{h(\tau)}{2}} \\
&\times \left(\varphi'^2 - \tanh'\frac{h(\tau)}{2}\left(\frac{i\varepsilon K}{\alpha_S}(\varphi'(\tau) + i\varepsilon h'(\tau)) + \frac{\Delta\mu}{2^{2\Delta-1}\alpha_S\cosh^{2\Delta}\left(\frac{h_L(\tau)-h_R(\tau)}{2}\right)}[h'(\tau)]^{2\Delta-1}\right)\right),
\end{aligned} \quad (91)$$

where we have introduced the approximation $B^\Delta = N$. The first term in the Lagrangian emerges from multiplier and denotes the flow equations, while the second term enforces the physical behaviors back to the Hermitian undeformed theory as $\lambda \to 0$. Since the coefficient C with N dependence has no impact on first term with deformation $\lambda$, the $T\bar{T}$ deformation in low energy limit vanishes when N is sufficiently large. We also note that the particle action in nonrelativistic limit should return to the undeformed form,

$$\frac{S}{N} = \int d\tau [\varphi'^2 - e^{2\varphi} + \mu e^{\varphi/2}], \quad (92)$$

with the constraint $\varphi = \log h'$.

## VII. CONCLUSION

In this work, we have investigated a $T\bar{T}$ deformation on both Hermitian and non-Hermitian two coupled SYK model.

We elaborated on the coupled SYK on a deformed entanglement property with a thermal field double and possible overlap between ground state and TFD, and it is not broken under $T\bar{T}$ deformation. In the thermal limit, additional correlations were established as a wormhole. We found that the deformation of large N theory could be reparametrized and show the same result as a shift of the model coupling constants $J$, $\mu$, and $\kappa$, as shown in Figs. 1–3. The Hawking-Page-like phase transition shifts towards gapless black hole and requires less coupling for traversable wormhole. The non-Hermiticity only appears in interaction terms with $\mu$ and breaks the symmetry between left and right models. It also leads to an increasing effect on free energy in wormhole phase. The free energy under deformation also undergoes a shift in the black hole phase entropy. $T\bar{T}$ deformation on phase structure is not influenced by non-Hermiticity. Additionally, we investigate the Lorentzian-time Green's function in critical points between black hole and wormhole phase. The wormhole solutions represented by real time Green's function gradually transform into black hole as the deformation parameter increases.

We have considered $NAdS_2$ spacetime with both Hermitian and non-Hermitian interaction operators, which only act on boundaries. And non-Hermitian effects break the symmetry of entanglement wormhole between left and right copies. These time-dependent boundary interactions have no dependence on finite-radial spacetime, and the form of corresponding $T\bar{T}$ flows in spacetime will not change. Energy-momentum flow and its trigger, the $T\bar{T}$ operator on cutoff radial sphere should be in the same form as $T\bar{T}$ deformation. In low energy limit, we obtain the deformed effective Schwarzian form for both dilaton geometry on the gravity side and the correlation function in the quantum model and consider its nonrelativistic limit.

## ACKNOWLEDGMENTS

We would like to thank Sizheng Cao, Xian-Hui Ge, and Song He for valuable suggestions. This work is supported by NSFC China (Grants No. 12275166, No. 11805117, and No. 11875184).

[1] J. Maldacena and D. Stanford, Remarks on the Sachdev-Ye-Kitaev model, Phys. Rev. D **94,** 106002 (2016).

[2] S. Sachdev and J. W. Ye, Gapless spin-fluid ground state in a random quantum Heisenberg magnet, Phys. Rev. Lett. **70,** 3339 (1993).

[3] S. Sachdev, Bekenstein-Hawking entropy and strange metals, Phys. Rev. X **5,** 041025 (2015).

[4] A. Kitaev, A simple model of quantum holography (2015), KITP Strings Seminar and Entanglement Program.

[5] J. Maldacena, D. Stanford, and Z. Yang, Conformal symmetry and its breaking in two-dimensional nearly anti-de Sitter space, Prog. Theor. Exp. Phys. **2016,** 12C104 (2016).

[6] J. Maldacena and X. L. Qi, Eternal traversable wormhole, arXiv:1804.00491.

[7] T. G. Zhou, L. Pan, Y. Chen, P. Zhang, and H. Zhai, Disconnecting a traversable wormhole: Universal quench dynamics in random spin models, Phys. Rev. Res. **3,** L022024 (2021).

[8] Y. Chen, X. L. Qi, and P. Zhang, Replica wormhole and information retrieval in the SYK model coupled to Majorana chains, J. High Energy Phys. 06 (2020) 121.

[9] X. L. Qi and P. Zhang, The coupled SYK model at finite temperature, J. High Energy Phys. 05 (2020) 129.

[10] T. G. Zhou and P. Zhang, Tunneling through an eternal traversable wormhole, Phys. Rev. B **102,** 224305 (2020).

[11] C. M. Bender, Making sense of non-Hermitian Hamiltonians, Rep. Prog. Phys. **70,** 947 (2007).

[12] C. M. Bender and S. Boettcher, Real spectra in non-Hermitian Hamiltonians having $PT$ symmetry, Phys. Rev. Lett. **80,** 5243 (1998).

[13] C. M. Bender, S. Boettcher, H. F. Jones, and P. N. Meisinger, Double-scaling limit of a broken symmetry quantum field theory in dimensions $D < 2$, J. Math. Phys. (N.Y.) **42,** 1960 (2001).

[14] C. M. Bender, S. Boettcher, H. F. Jones, P. N. Meisinger, and M. S. Simsek, Bound states of non-Hermitian quantum field theories, Phys. Lett. A **291,** 197 (2001).

[15] A. M. García-García and V. Godet, Euclidean wormhole in the Sachdev-Ye-Kitaev model, Phys. Rev. D **103,** 046014 (2021).

[16] A. M. García-García, V. Godet, C. Yin, and J. P. Zheng, Euclidean-to-Lorentzian wormhole transition and gravitational symmetry breaking in the Sachdev-Ye-Kitaev model, Phys. Rev. D **106,** 046008 (2022).

[17] W. Cai, S. Cao, X. H. Ge, M. Matsumoto, and S. Sin, Non-Hermitian quantum system generated from two coupled Sachdev-Ye-Kitaev models, Phys. Rev. D **106,** 106010 (2022).

[18] S. He and Z. Y. Xian, $T\bar{T}$ deformation on multiquantum mechanics and regenesis, Phys. Rev. D **106,** 046002 (2022).

[19] S. He, P. Hang, C. Lau, Z. Y. Xian, and L. Zhao, Quantum chaos, scrambling and operator growth in $T\bar{T}$ deformed SYK models, J. High Energy Phys. 12 (2022) 070.

[20] F. A. Smirnov and A. B. Zamolodchikov, On space of integrable quantum field theories, Nucl. Phys. **B915,** 363 (2017).

[21] A. B. Zamolodchikov, Expectation value of composite field $T\bar{T}$ in two-dimensional quantum field theory, arXiv:hep-th/0401146.

[22] A. Cavagli, Stefano Negro, Istv'an M. Sz'ecs'enyi, and Roberto Tateo, $T\bar{T}$-deformed 2D quantum field theories, J. High Energy Phys. 10 (2016) 112.

[23] D. J. Gross, J. Kruthoff, A. Rolph, and E. Shaghoulian, $T\bar{T}$ in $AdS_2$ and quantum mechanics, Phys. Rev. D **101,** 026011 (2020).

[24] D. J. Gross, J. Kruthoff, A. Rolph, and E. Shaghoulian, Hamiltonian deformations in quantum mechanics, $T\bar{T}$ and SYK, Phys. Rev. D **102,** 046019 (2020).

[25] L. Castillejo, R. H. Dalitz, and F. J. Dyson, Low scattering equation for the charged and neutral scalar theories, Phys. Rev. **101,** 453 (1956).

[26] S. Dubovsky, R. Flauger, and V. Gorbenko, Solving the simplest theory of quantum gravity, J. High Energy Phys. 09 (2012) 133.

[27] R. Conti, L. Iannella, S. Negro, and R. Tateo, Generalised Born-Infeld models, Lax operators and the $T\bar{T}$ perturbation, J. High Energy Phys. 11 (2018) 007.

[28] R. Conti, S. Negro, and R. Tateo, The $T\bar{T}$ perturbation and its geometric interpretation, J. High Energy Phys. 02 (2019) 085.

[29] G. Hernandez-Chifflet, S. Negro, and A. Sfondrini, Flow equations for generalized $T\bar{T}$ deformations, Phys. Rev. Lett. **124,** 200601 (2020).

[30] O. Aharony, S. Datta, A. Giveon, Y. Jiang, and D. Kutasov, Modular invariance and uniqueness of $T\bar{T}$ deformed CFT, J. High Energy Phys. 01 (**2019**) 086.

[31] L. Susskind, Entanglement and chaos in de Sitter holography: An SYK example, J. Hologr. Appl. Phys. **1**, 1 (2021).

[32] G. Bonelli, N. Doroud, and M. Zhu, $T\bar{T}$-deformations in closed form, J. High Energy Phys. 06 (2018) 149.

[33] O. Aharony and Z. Komargodski, The effective theory of long strings, J. High Energy Phys. 05 (2013) 118.

[34] M. Taylor, TT deformations in general dimensions, Adv. Theor. Math. Phys. **27**, 37 (2023).

[35] R. Conti, S. Negro, and R. Tateo, Conserved currents and $T\bar{T}$s irrelevant deformations of 2D integrable field theories, J. High Energy Phys. 11 (2019) 120.

[36] A. B. Zamolodchikov, Expectation value of composite field $T\bar{T}$ in two-dimensional quantum field theory, arXiv:hep-th/0401146.

[37] L. McGough, M. Mezei, and H. Verlinde, Moving the CFT into the bulk with $T\bar{T}$, J. High Energy Phys. 04 (2018) 010.

[38] T. Hartman, J. Kruthoff, E. Shaghoulian, and A. Tajdini, Holography at finite cutoff with a $T^2$ deformation, J. High Energy Phys. 03 (2019) 004.

[39] P. Caputa, S. Datta, and V. Shyam, Sphere partition functions and cut-off AdS, J. High Energy Phys. 05 (2019) 112.

[40] V. Shyam, Background independent holographic dual to $T\bar{T}$ deformed CFT with large central charge in 2 dimensions, J. High Energy Phys. 10 (**2017**) 108.

[41] S. Dubovsky, V. Gorbenko, and G. Hernandez-Chifflet, $T\bar{T}$ partition function from topo logical gravity, J. High Energy Phys. 09 (2018) 158.

[42] D. Das, S. Sridip, and A. Sarkar, Wormholes and half wormholes under irrelevant deformation, Phys. Rev. D **106**, 066014 (2022).

[43] S. Sahoo, E. Lantagne-Hurtubise, S. Plugge, and M. Franz, Traversable wormhole and Hawking-Page transition in coupled complex SYK models, Phys. Rev. Res. **2**, 043049 (2020).

[44] S. Cao, Y. C. Rui, and X. H. Ge, Thermodynamic phase structure of complex Sachdev-Ye-Kitaev model and charged black hole in deformed JT gravity, arXiv:2103.16270.

[45] A. S. Matsoukas-Roubeas, F. Roccati, J. Cornelius, Z. Xu, A. Chenu, and A. del Campo, Non-Hermitian Hamiltonian deformations in quantum mechanics, J. High Energy Phys. 01 (2023) 060.

[46] J. Cornelius, Z. Xu, A. Saxena, A. Chenu, and A. del Campo, Spectral filtering induced by non-Hermitian evolution with balanced gain and loss: Enhancing quantum chaos, Phys. Rev. Lett. **128**, 190402 (2022).

[47] T. Nosaka and T. Numasawa, Chaos exponents of SYK traversable wormholes, J. High Energy Phys. 02 (2021) 150.

[48] S. Plugge, E. Lantange-Hurtubise, and M. Franz, Revival dynamics in a traversable wormhole, Phys. Rev. Lett. **124**, 221601 (2020).

[49] P. Kakashvili and C. J. Bolech, Time-loop formalism for irreversible quantum problems: Steady-state transport in junctions with asymmetric dynamics, Phys. Rev. B **78**, 033103 (2008).

[50] A. Almheiri and J. Polchinski, Models of $AdS_2$ backreaction and holography, J. High Energy Phys. 11 (2015) 014.

[51] R. Jackiw, Lower dimensional gravity, Nucl. Phys. **B252**, 343 (1985).

[52] C. Teitelboim, Gravitation and Hamiltonian structure in two space-time dimensions, Phys. Lett. **126B**, 41 (1983).

[53] J. Maldacena, D. Stanford, and Z. Yang, Conformal symmetry and its breaking in two dimensional nearly anti-de-Sitter space, Prog. Theor. Exp. Phys. **2016**, 12C104 (2016).

[54] T. Hartman, J. Kruthoff, E. Shaghoulian, and A. Tajdini, Holography at finite cutoff with a $T^2$ deformation, J. High Energy Phys. 03 (2019) 004.

[55] C. De Rham and A. Matas, Ostrogradsky in theories with multiple fields, J. Cosmol. Astropart. Phys. 06 (2016) 041.

[56] W. Cai, X. H. Ge, and G. H. Yang, Diffusion in higher dimensional SYK model with complex fermions, J. High Energy Phys. 01 (2018) 076.

[57] R. A. Davison, W. Fu, A. Georges, Y. Gu, K. Jensen, and S. Sachdev, Thermoelectric transport in disordered metals without quasiparticles: The Sachdev-Ye-Kitaev models and holography, Phys. Rev. B **95**, 155131 (2017).